\documentclass[pdflatex,sn-mathphys-num]{sn-jnl}

\usepackage{graphicx}%
\usepackage{multirow}%
\usepackage{amsmath,amssymb,amsfonts}%
\usepackage{amsthm}%
\usepackage{mathrsfs}%
\usepackage[title]{appendix}%
\usepackage{xcolor}%
\usepackage{textcomp}%
\usepackage{manyfoot}%
\usepackage{booktabs}%
\usepackage{hyphenat}
\usepackage{algorithm}%
\usepackage{algorithmicx}%
\usepackage{algpseudocode}%
\usepackage{listings}%

\theoremstyle{thmstyleone}%
\theoremstyle{thmstyletwo}%

\theoremstyle{thmstylethree}%

\begin{document}

\title[Article Title]{The Atomic Bomb: Its History and the Struggles of Scientists}


\author*[1,2]{\fnm{Shoji} \sur{Nagamiya}}\email{shoji.nagamiya@riken.jp}


\affil[1]{\orgdiv{RIKEN},  \orgaddress{\street{2-1 Hirosawa}, \city{Wako-shi}, \postcode{351-0106}, \state{Saitama-ken}, \country{Japan}}}

\affil*[2]{\orgdiv{KEK},  \orgaddress{\street{1-1 Oho}, \city{Tsukuba-shi}, \postcode{305-0801}, \state{Ibataki-ken}, \country{Japan}}}

\abstract{
In this article, I trace the early historical developments that ultimately led to the creation of the atomic bomb. Even after the weapon’s completion, a number of scientists continued to argue that nuclear armaments were indispensable for maintaining the global balance of political power [1]. This study focuses on several scientists who confronted profound moral dilemmas concerning the bomb’s use against Japan. Some openly opposed its deployment, others sought to warn a Japanese scientist in the hope of averting further devastation, and still others expressed deep remorse in its aftermath. In addition, the experience of an individual directly affected by the bombing is discussed.

By examining these episodes, this article aims to contribute to the ongoing discourse on how scientific research should be guided by ethical principles in the future.
}

\keywords{Atomic Bomb, Its History, Dilemma among Scientists}

\maketitle

\section{Preface}\label{sec1}

In the spring of~2024, a film about J.~Robert~Oppenheimer attracted considerable public attention in Japan. I saw it only once, yet I noticed that it largely omitted the progress of atomic-bomb research that preceded Oppenheimer’s involvement. Conversely, the appearance of figures such as Edward~Teller and I.~I.~Rabi in the film evoked personal memories, as I had spoken with both of them. I also learned certain aspects from the movie that had previously escaped my attention.

Beginning in~1973, I spent nearly nine years at the University of California, Berkeley, where Oppenheimer had once worked, and later about ten and a half years at Columbia University in New~York, where Rabi had been. Reflecting on what I learned during those years about the history of the atomic bomb, I would like to share some background that was not depicted in the \textit{Oppenheimer} film. Even among the scientists directly involved in the bomb’s development and use, there existed profound internal conflicts and moral struggles.

Meanwhile, in the autumn of~2024, encouraging news emerged that the Japan~Confederation~of~A-~and~H-Bomb~Sufferers~Organisations (Hidankyo) had been nominated for the Nobel~Peace~Prize. The suffering of those who experienced the atomic bombings will, of course, never fade. I deeply admire the tireless efforts of Hidankyo, and in this article I also wish to reflect on a small part of the pain endured by the survivors.

\section{Fermi and the Neutron Beam}\label{sec2}
\begin{figure}[htbp]
\centering
\includegraphics[width=0.8\textwidth]{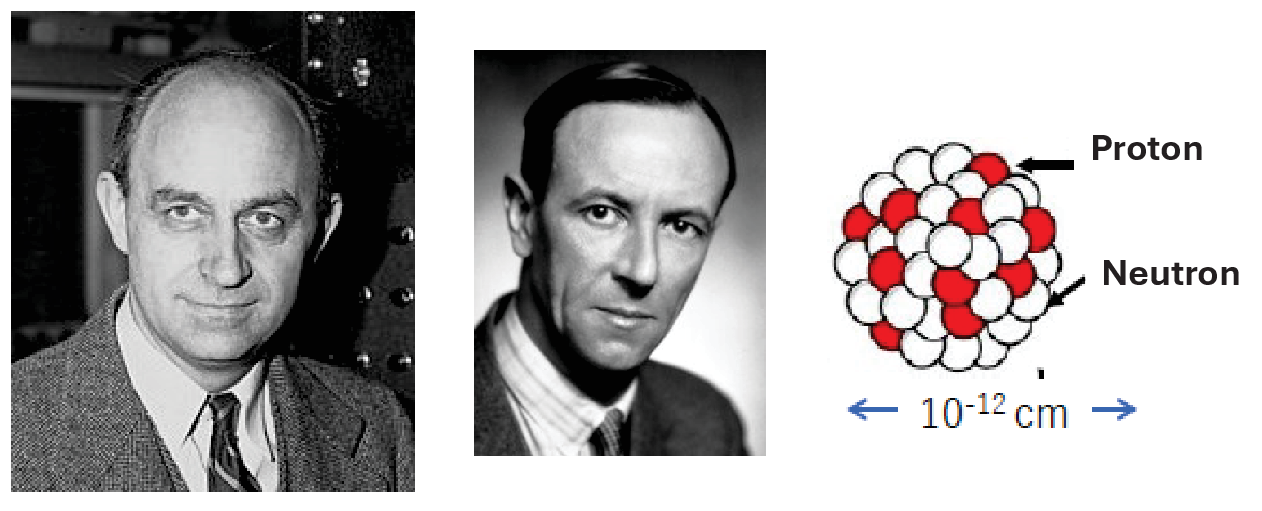}
\caption{Left: Enrico~Fermi (1901--1954). Right: James~Chadwick and a schematic illustration of the atomic nucleus, showing its constituent protons and neutrons.}
\label{fig1}
\end{figure}

Anyone who has studied physics is familiar with the name Enrico~Fermi, shown in Fig.~\ref{fig1}~(left). This story, however, reaches back many years. When I was a student in~1966, our physics department held an event called the \textit{Newton~Festival} around Christmastime. On that occasion, we students conducted a small survey among our professors, asking, ``Who is the greatest figure in the history of physics?'' The responses did not name Einstein or any other widely celebrated physicist, but rather Fermi. Such was his reputation---renowned for his exceptionally broad contributions to virtually every branch of physics, both theoretical and experimental.

Matter consists of atoms and molecules. Each atom contains a positively charged nucleus at its centre, around which light, negatively charged electrons revolve. To visualise the scale, if an atom were the size of a baseball stadium, its nucleus would be smaller than a baseball---yet this minute core accounts for more than 99~\% of the atom’s total mass.

In the early twentieth century, physicists were confronted with two fundamental questions:\\
\noindent (a)~What particles constitute the atomic nucleus?\\
\noindent (b)~Is it possible to create new nuclei artificially?

To address these problems, John~Cockcroft and Ernest~Walton in the United~Kingdom constructed a particle accelerator in~1932. Although this accelerator succeeded in inducing a nuclear reaction, its capability was limited. In the United~States, Ernest~Lawrence developed a different type of accelerator---the cyclotron in 1932. Using this device, Emilio~Segr\`e and Carlo~Perrier later discovered the first artificially produced element, technetium, in~1937~\cite{seg}.

Amid these developments, in~1932, James~Chadwick in the United~Kingdom confirmed the existence of the neutron. 
As illustrated in Fig.~\ref{fig1}~(right), the neutron has nearly the same mass as the proton, 
and it soon became evident that the atomic nucleus is composed of positively charged protons 
and electrically neutral neutrons. 
It was further established that negatively charged electrons
orbit the nucleus in numbers equal to those of the protons, thereby completing the
current understanding of atomic structure.

At that time, most researchers focused on accelerating charged particles, such as protons or atomic nuclei, using electric fields. Enrico~Fermi, however, turned his attention to beams of neutrons, which---being electrically neutral---were extremely difficult to control experimentally. Despite these challenges, Fermi conducted a series of pioneering experiments with neutrons and succeeded in producing new isotopes. For this achievement, described as ``nuclear reactions produced by neutron bombardment,'' he was awarded the~1938~Nobel~Prize in~Physics, only six years after the neutron’s discovery.

As noted at the beginning of this article, nearly every physicist is familiar with Fermi’s name, yet surprisingly few are aware that this specific work formed the basis of his Nobel recognition---perhaps because his contributions spanned such a wide range of fields within physics. According to Fermi’s Nobel~Lecture~\cite{fermi}, he and his collaborators initially believed that they had synthesised elements~93 and~94 in Rome. This interpretation, however, was later shown to be incorrect~\cite{fermi1}. In~1940, element~93~(neptunium, Np) and element~94~(plutonium, Pu) were discovered through cyclotron experiments at Berkeley by E.~McMillan~\textit{et~al.}~\cite{mcm} and G.~Seaborg~\textit{et~al.}~\cite{sea}, respectively.

Following Fermi’s pioneering publications, Otto~Hahn and Lise~Meitner in Germany, as well as Ir\`ene~Curie and Fr\'ed\'eric~Joliot in France, undertook systematic studies of
new radioactive isotopes and neutron‐induced reactions~\cite{stu}. Of course, no one at the time could have foreseen that these neutron-induced experiments would eventually open the path toward the development of the atomic bomb.

\section{Fermi’s Escape to the United States}\label{sec3}

In~1938, the Italian physicist Enrico~Fermi faced increasingly difficult circumstances as Italy came under Mussolini’s fascist regime and entered into alliance with Nazi~Germany. During the summer and autumn of that year, the so-called \textit{Racial~Protection~Laws} were enacted, declaring that only Aryans were ``true Italians'' and---much as in Germany---imposing severe discrimination against Jews, particularly within universities and research institutions.

\begin{figure}[htbp]
\centering
\includegraphics[width=1.0\textwidth]{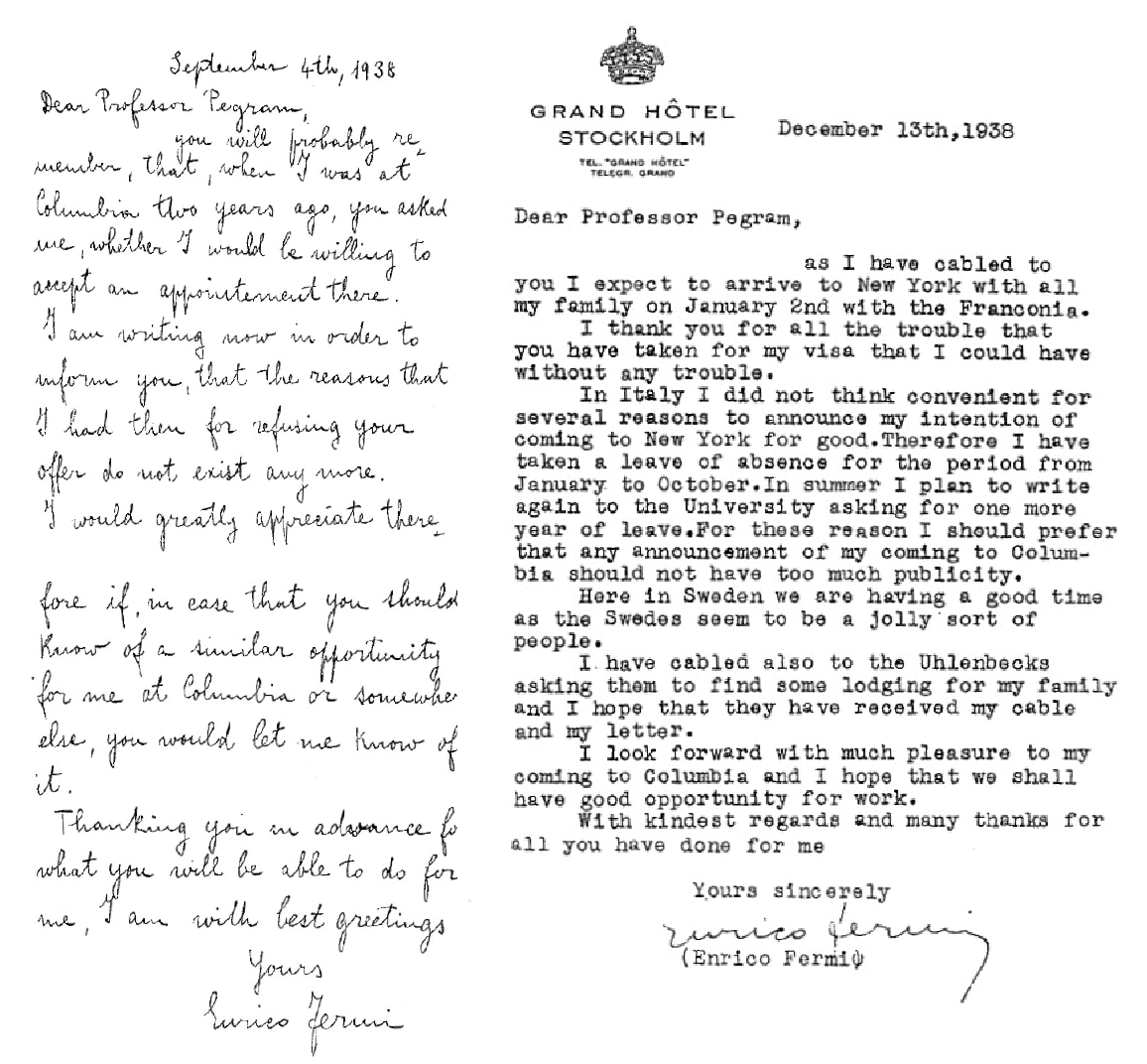}
\caption{Two letters from Enrico~Fermi to G.~B.~Pegram, Chair of the Department of Physics at Columbia~University. The letter on the left, dated~4~September~1938, inquires about a possible position at Columbia, while the one on the right, written shortly thereafter, informs Pegram of Fermi’s plan to travel from Stockholm to the United~States on the occasion of the Nobel~Prize ceremony. Both letters are reproduced from Ref.~\cite{lee}.}
\label{fig2}
\end{figure}

On~4~September~1938, shortly after these laws were promulgated, Fermi wrote to G.~B.~Pegram, then Chair of the Department of Physics at Columbia~University. As shown in Fig.~\ref{fig2}~(left), he asked:
``Regarding the position you offered me two years ago, even if it is not the same one, might it still be available?'  His motivation was both personal and professional, as his wife, Laura~Fermi, was Jewish. Pegram promptly arranged a professorship for him at Columbia~University.

Several decades later, in the~1990s, when I was serving as Chair of the same department, we discovered a large collection of historical correspondence while cleaning out the departmental basement. These letters were later compiled and published by T.~D.~Lee~\cite{lee}, whose office was located diagonally across from mine. Among them were letters---mostly handwritten---from H.~A.~Lorentz, Albert~Einstein, Niels~Bohr, R.~A.~Millikan, two from Max~Planck, and four from Enrico~Fermi. One of these---the very letter to Pegram mentioned above---is reproduced in Fig.~\ref{fig2}~(left).

Soon after writing to Pegram, Fermi was awarded the Nobel~Prize in~Physics. He and his family used their trip to Stockholm for the award ceremony as an opportunity to escape to the United~States, travelling directly from Sweden. One of the confidential letters related to this escape is reproduced in Fig.~\ref{fig2}~(right). By the beginning of January~1939, Fermi had assumed his new position as Professor of Physics at Columbia~University.

\section{From the Discovery of Nuclear Fission to the Chain Reaction}\label{sec4}

\begin{figure}[htbp]
\centering
\includegraphics[width=0.8\textwidth]{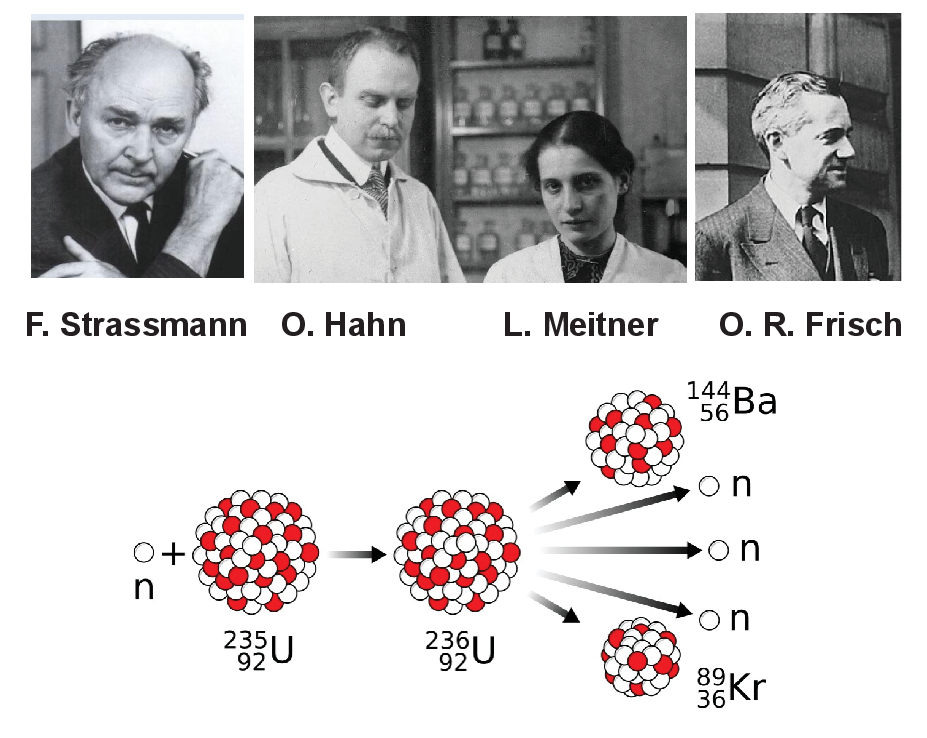}
\caption{Scientists associated with the discovery of nuclear fission. From left: Friedrich~Strassmann, Otto~Hahn, Lise~Meitner, and Otto~R.~Frisch~(top). Their schematic explanation of the fission process is shown at the bottom.}
\label{fig3}
\end{figure}

Meanwhile, in Europe, the neutron-beam experiments initiated by Enrico~Fermi continued---particularly in Germany, where researchers were also irradiating uranium~(U) with neutrons. Toward the end of~1938, the German chemists Otto~Hahn and Friedrich~Strassmann detected barium~(Ba)---an element of considerably lower atomic mass---among the reaction products. Since uranium is element~92 and barium~56, the discrepancy in atomic number was too large to be ignored. They promptly informed their collaborator Lise~Meitner, who had by then fled to Sweden. Meitner, together with her nephew Otto~Frisch, analysed the results in January 1939 and concluded that 
that the uranium nucleus had split into two lighter nuclei
---principally Ba and krypton~(Kr). By analogy with biological ``cell fission,'' they named this process \textit{nuclear fission}. 
Initially they assumed the entire uranium nucleus was susceptible, but Niels Bohr emphasized that only the rare isotope  ${}^{235}$U  undergoes fission with slow neutrons, whereas  ${}^{238}$U  does not  (see Fig.~\ref{fig3}).

To explain further, when a neutron is incident on uranium-235 (which constitutes about~0.7\% of natural uranium), the nucleus splits into barium-144, krypton-89, and releases three neutrons. Denoting a neutron as~$n$, the reaction can be expressed as
\begin{equation}
n + {}^{235}\mathrm{U} \rightarrow {}^{144}\mathrm{Ba} + {}^{89}\mathrm{Kr} + 3n.
\end{equation}
According to Einstein’s principle that mass difference is equivalent to energy, the difference between the left- and right-hand sides of the equation corresponds to roughly~200~MeV. Since the combustion of paper releases only a few electron volts per molecule, this process yields nearly one hundred million times more energy. Later studies revealed that the above channel is not unique: there exist dozens of possible fission products, and the average number of neutrons emitted per fission of~${}^{235}$U was found to be about~2.5.

In~January~1939, shortly after Fermi’s arrival in the United~States, Niels~Bohr gave a lecture at Princeton~University announcing this discovery. Although it was Willis~Lamb who actually attended the talk, Fermi soon learned of it indirectly. He immediately recognised the process as nuclear fission and, in~March~1939---shortly after assuming his position at Columbia~University in New~York---constructed a small uranium pile to verify fission in~${}^{235}$U. (According to historical accounts, Herbert~Anderson began preliminary experiments immediately after word of the discovery leaked out---before Fermi had even started building his first pile~\cite{reed1}.)

\begin{figure}[htbp]
\centering
\includegraphics[width=0.95\textwidth]{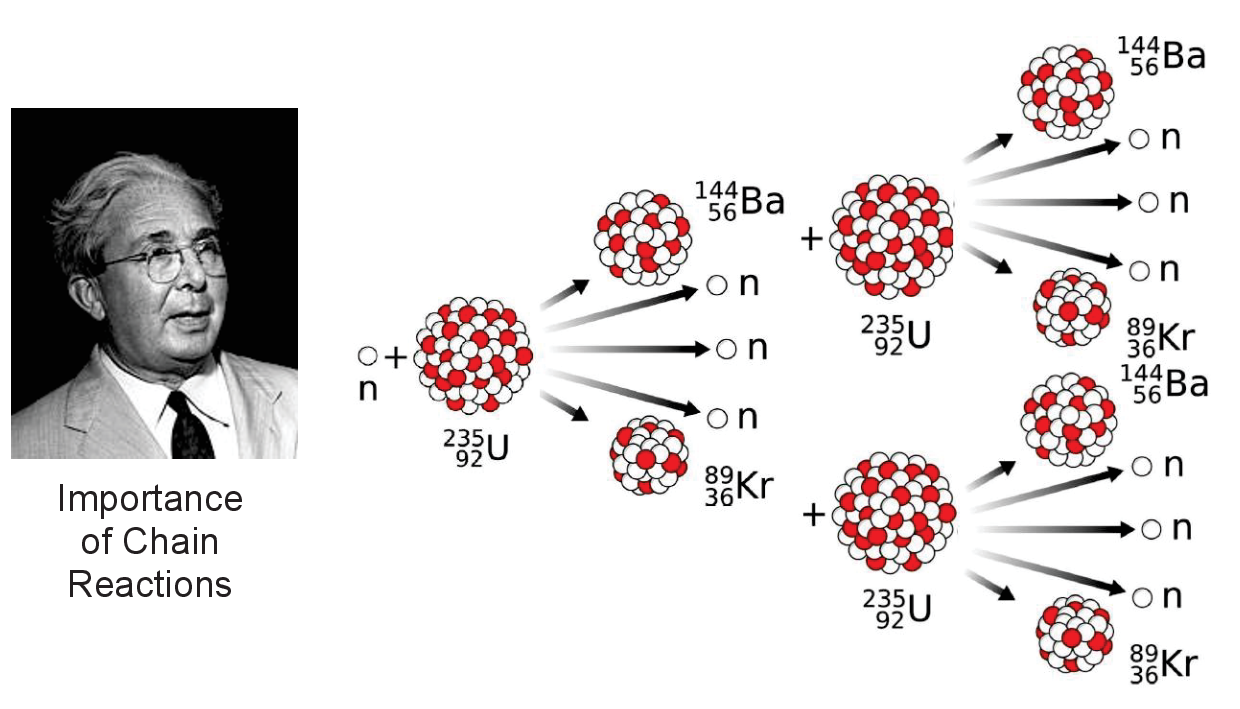}
\caption{Le\'o~Szil\'ard~(left) and his concept of a self-sustaining nuclear chain reaction capable of releasing an enormous amount of energy.}
\label{fig4}
\end{figure}

Around the same time, the Hungarian-born Jewish physicist Le\'o~Szil\'ard (see Fig.~\ref{fig4}, left) was also at Columbia~University. Szil\'ard and Enrico~Fermi would soon collaborate, but even before that, Szil\'ard had begun theorising that if nuclear fission released two or three neutrons, and those neutrons in turn struck additional~${}^{235}$U nuclei, further fission reactions could occur (see Fig.~\ref{fig4}, right). Assuming that each fission emitted, on average, about~2.5~neutrons, the second generation would produce roughly six neutrons, and the energy released would grow exponentially. If such chain reactions could be sustained, they would result in an enormous release of energy.

Szil\'ard had, in fact, conceived the idea of a nuclear chain reaction several years earlier---in~1933---while crossing Russell~Square in~London. He even filed a British patent (No.~630,726; 1934) describing the concept of an ``atomic bomb.'' Although no such reaction had yet been observed experimentally, his insight clearly anticipated later developments.

\section{Toward CP-1 (Chicago Pile No.1)}\label{sec5}

Recognizing the far-reaching implications of neutron-induced chain reactions, Szil\'ard drafted a letter to President Franklin D.~Roosevelt and persuaded Albert Einstein to sign it.  The letter, dated 2~August~1939, warned of the possibility that Nazi Germany might attempt to construct extremely powerful bombs.  Shortly after the letter was dispatched, Germany invaded Poland on 1~September~1939, marking the beginning of the Second World War.  The United States, however, remained officially neutral.

Even if the weaponization of fission were to be pursued, it was believed at the time that several tons of natural uranium would be required, raising serious logistical concerns.  In response to these issues, Roosevelt authorized further investigation by establishing the Advisory Committee on Uranium, which in October~1939 allocated an initial sum of \$6,000 for the purchase of uranium
and graphite.  This modest grant marked the beginning of sustained U.S.\ government support for nuclear research~\cite{lee}.

Among those who subsequently became involved were James Conant, President of Harvard University, and Arthur Compton, who would later assume a central role at the Metallurgical Laboratory of the University of Chicago and in the development of CP--1.  Fermi and Szil\'ard initially began constructing experimental piles at Columbia University; several subcritical assemblies were built there in 1940--41~\cite{reed1}.  It was not the impracticality of New York as a site but rather Compton's administrative decision, taken in early 1942, to centralize all pile work in Chicago that led to the relocation of the programme.

\begin{figure}[htbp]
\centering
\includegraphics[width=1.0\textwidth]{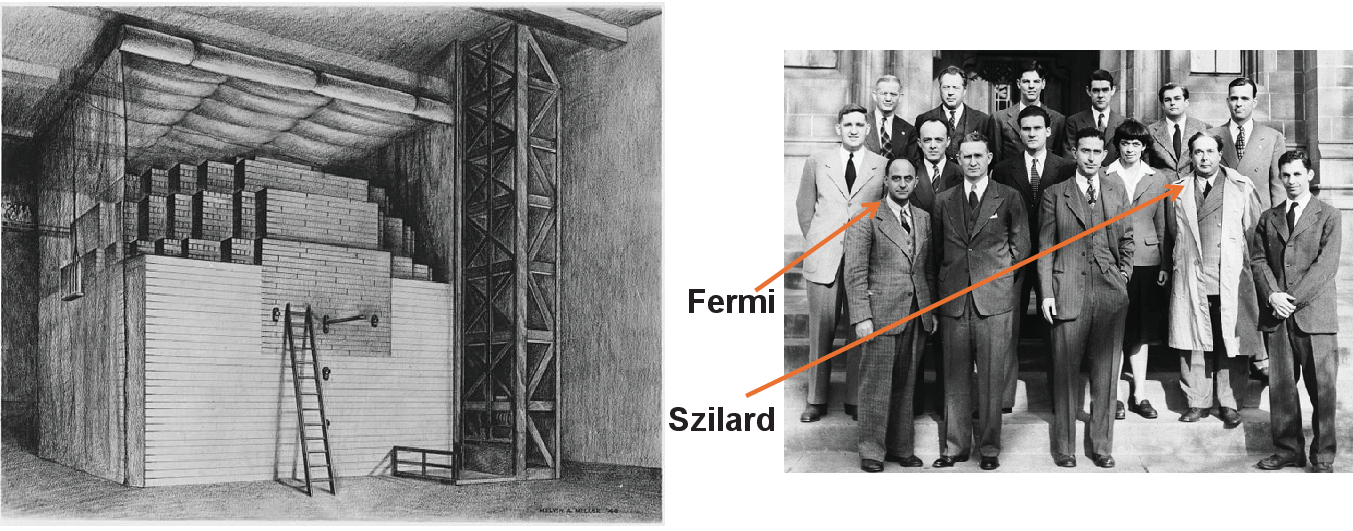}
\caption{The Chicago~Pile~No.~1~(CP-1) reactor directed by Enrico~Fermi~(left) and several of the scientists involved in its construction~(right).}
\label{fig5}
\end{figure}

Skipping the detailed sequence of experiments, the Chicago group eventually assembled approximately 42~t of natural uranium (a mixture of uranium oxide and metal) embedded within about 330~t of graphite blocks, together with cadmium control rods.  After roughly three years of reactor research, the construction
of Chicago Pile No.\,1 was completed in late November~1942, and on 2~December~1942 the world’s first self-sustaining nuclear chain reaction was successfully achieved (see Fig.~\ref{fig5}).

Upon receiving the result by telephone, Arthur Compton informed James Conant in coded language so as not to disclose sensitive information over the line:  
\begin{quote}
``The Italian navigator has landed in the New World.  The natives were very friendly.''  
\end{quote}
The phrase, of course, referred to Fermi and to the successful achievement of controlled criticality.

Although the CP--1 experiment remained an outstanding achievement in reactor physics, it was already embedded within the emerging Manhattan Project.  Given that natural uranium contains only about 0.7\% of ${}^{235}$U and more than 99\%
of ${}^{238}$U, an inevitable consequence of sustained neutron irradiation was the production of plutonium, first identified by Glenn T.~Seaborg somewhat earlier in 1940.  In the months following CP--1, ${}^{239}$Pu rapidly became, together with ${}^{235}$U, one of the two principal fissile materials of the
Manhattan Project.

\section{Developments in the United Kingdom}\label{sec6}

\begin{figure}[htbp]
\centering
\includegraphics[width=0.8\textwidth]{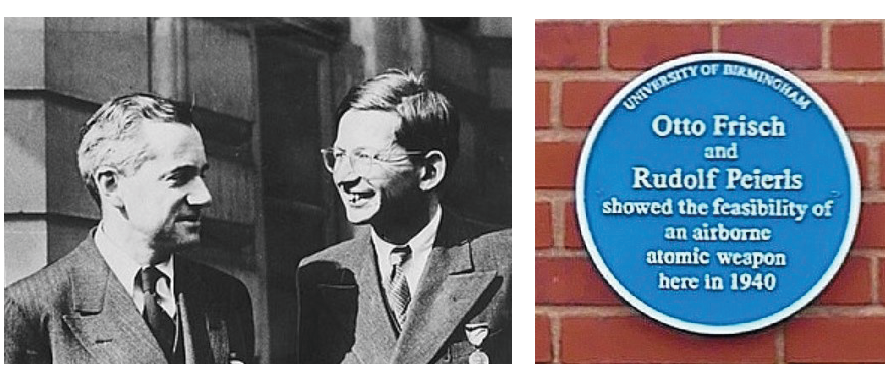}
\caption{Otto~Frisch and Rudolf~Peierls proposed the possibility of constructing an atomic bomb using enriched~${}^{235}$U~(left). The commemorative plaque at the University~of~Birmingham~(right) marks their~1940 proposal for an air-transportable atomic bomb.}
\label{fig6}
\end{figure}

We now turn to developments in the United Kingdom around 1939.  Otto Frisch, an Austrian-born Jewish refugee who had contributed decisively to the discovery of nuclear fission, was working in Birmingham with Rudolf Peierls, a German-born Jewish refugee who would later become a leading figure in condensed matter physics (see Fig.~\ref{fig6}).  

Building on ideas originally proposed by Szil\'ard and on experimental results obtained by Joliot-Curie and collaborators, Frisch and Peierls considered the possibility of using highly enriched ${}^{235}$U as the basis of an exceptionally powerful weapon.  They showed that if ${}^{235}$U were available in nearly pure form, only a few kilograms---small enough to be delivered by aircraft---would suffice for a nuclear explosion.  Their results were summarised in the celebrated \textit{Frisch--Peierls Memorandum} of March~1940~\cite{reed2}.

Prompted by this memorandum, the MAUD Committee was established in mid-1940 under the chairmanship of George P.~Thomson to examine uranium-enrichment methods, reactor design, the properties of ${}^{235}$U, and the theoretical basis of fission weapons.  The committee, which included several \'{e}migr\'e and French scientists, conducted intensive studies that would prove decisive for subsequent Allied programmes.

By mid-1941 the MAUD Committee completed its final reports, concluding that an atomic bomb could be constructed with approximately 11~kg of enriched ${}^{235}$U and recommending that such a project be pursued urgently, ideally
in collaboration with the United States.  Shortly thereafter, the British government launched a top-secret nuclear-weapons programme under the codename \textit{Tube Alloys}~\cite{tube}.  

The sustained German bombing of London started in September 1940 and the severe financial constraints of wartime Britain made large-scale enrichment facilities impractical on British soil, compelling increasing reliance on American industrial resources.
In July~1941 the MAUD reports were transmitted to Washington, where they had a profound impact on American assessments of the feasibility of an atomic bomb. At that time, the United States was still officially neutral, although providing substantial support to Britain through Lend--Lease.  The Japanese attack on Pearl Harbor on 7~December~1941 brought the United States formally into the war.  

Shortly thereafter, during the Arcadia Conference (22~December~1941--14~January~1942), Prime Minister Winston Churchill and President Franklin D.~Roosevelt reaffirmed close Anglo--American scientific cooperation, laying part of the political groundwork for what would eventually
become a joint atomic-bomb programme.

As noted above, on 2~December~1942 the CP--1 experiment at Chicago achieved the first self-sustaining chain reaction, thereby also demonstrating the feasibility of producing ${}^{239}$Pu on a large scale.  The design of such production reactors was undertaken by Eugene Wigner and his group at the Metallurgical Laboratory, with Alvin Weinberg playing a key role in the graphite-moderated reactor concepts~\cite{wein}.  Note that Wigner later became renowned for his pioneering contributions to the
theory of symmetry in quantum and particle physics.

Following these developments, ${}^{235}$U enrichment at Oak Ridge in Tennessee and ${}^{239}$Pu production at Hanford in Washington State---the latter using reactors constructed by the DuPont Company---progressed rapidly.  Together these two programmes became the principal industrial pillars of the Manhattan Project.  

\begin{figure}[htbp]
\centering
\includegraphics[width=1.0\textwidth]{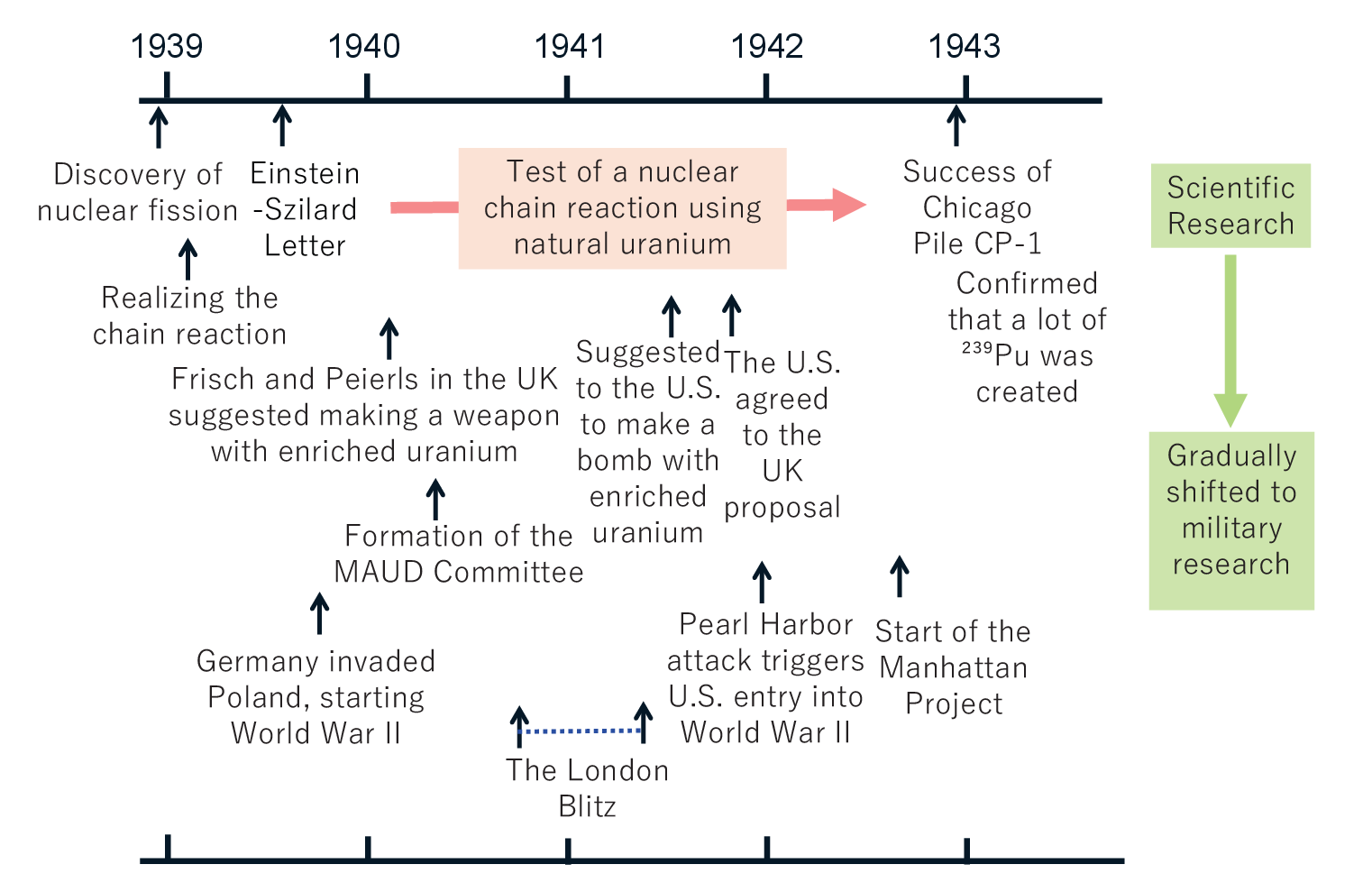}
\caption{Timeline illustrating major scientific and organisational developments during the four years preceding the initiation of the Manhattan~Project.}
\label{fig7}
\end{figure}

Fig.~\ref{fig7} summarises the major advances leading from the discovery of nuclear fission and, subsequently, the Frisch--Peierls Memorandum, to the completion of CP--1 and the initiation of full-scale wartime production.

\section{The Manhattan Project}\label{sec7}

The Manhattan Project was formally initiated in August~1942 under the administration of President Franklin~D.~Roosevelt. Many publications discuss the programme in depth~\cite{rho,reed3}; therefore, only a concise overview is given here.

\begin{figure}[htbp]
\centering
\includegraphics[width=1.0\textwidth]{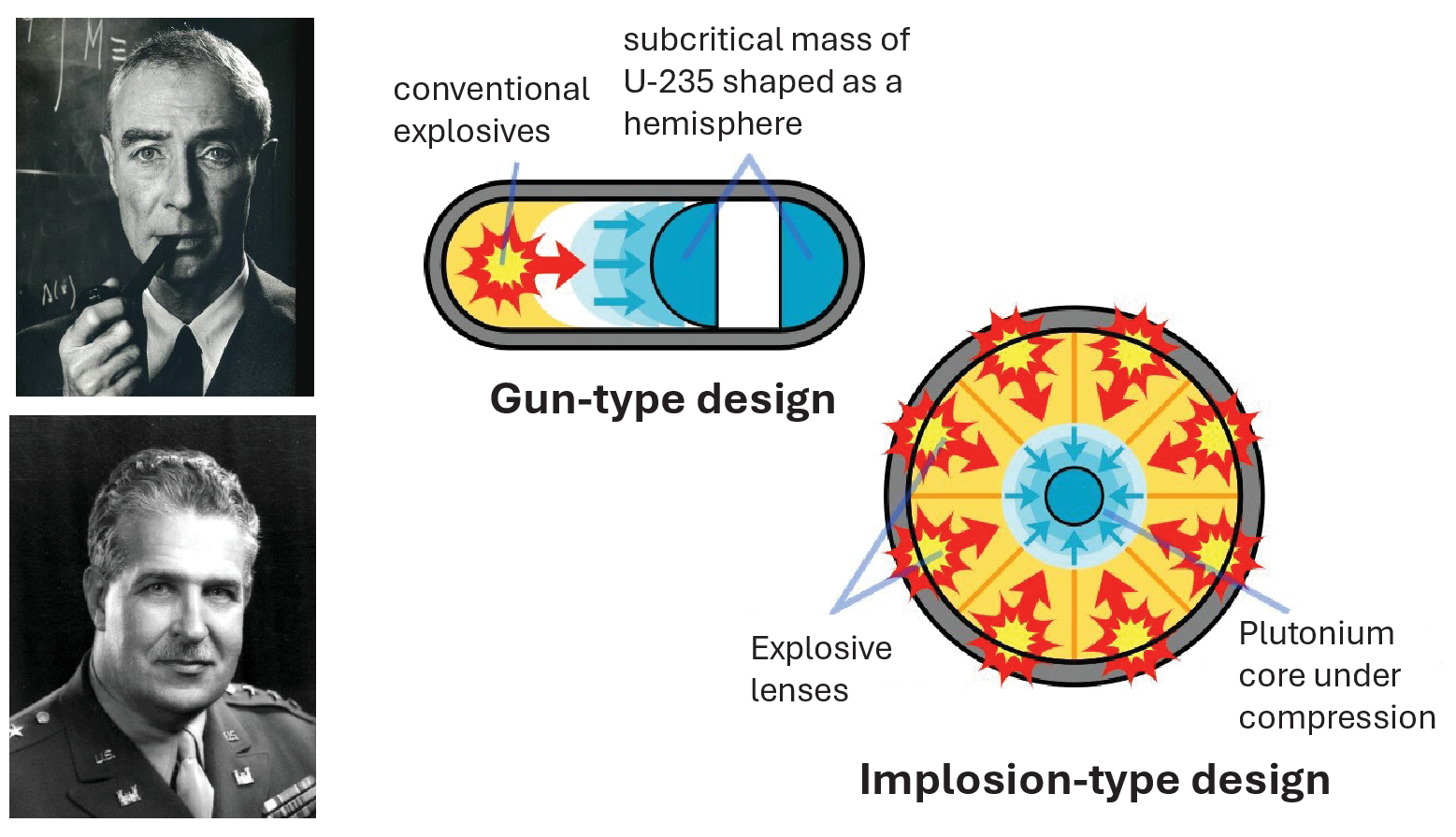}
\caption{Left: J.~Robert~Oppenheimer (top), scientific director of the Manhattan Project, and General~Leslie~R.~Groves (bottom), the overall project director and military commander. Right: Schematic diagrams of the gun-type bomb (``Little Boy'') and the implosion-type bomb (``Fat Man''); the latter acquired its nickname because contemporaries likened its bulky exterior to Winston~Churchill.}
\label{fig8}
\end{figure}

In September~1942, General Leslie~R.~Groves was appointed overall military director of the project, while J.~Robert Oppenheimer became its scientific director (see Fig.~\ref{fig8}). Although most participating scientists were American, a substantial number of British, Canadian, and European émigré scientists—many of them later naturalised in the United Kingdom or the United States—contributed significantly.

The design laboratory for nuclear weapons was established at Los Alamos, a remote mesa in New Mexico. A notable fraction of the scientists working on the project were Jewish refugees from Nazi persecution, and this background imparted a strong moral and emotional dimension to their efforts. At the same time, U.S.\ industrial capacity—unmatched at the time—was mobilised on an unprecedented scale. Oak Ridge, Tennessee, became the centre of uranium-enrichment activities. By the spring of 1945, weapons-grade levels of enrichment (typically 80--89\% $^{235}$U) had been achieved. Harold~C.~Urey oversaw development of the gaseous-diffusion method, while Ernest~O.~Lawrence supervised electromagnetic separation using calutrons at the Y-12 plant; both contributions were critical to securing sufficient enriched uranium.

The uranium weapon, ``Little Boy,'' employed a linear gun-type assembly (Fig.~\ref{fig8}, middle). When the cordite powder charge was ignited, it propelled a cylindrical uranium projectile into a stationary uranium target, forming a supercritical configuration and initiating a prompt neutron chain reaction. A tungsten tamper and neutron reflector surrounded the core to enhance neutron economy. Because only about 64~kg of enriched uranium were available, the device was never tested prior to combat use. Consequently, only about 1.38\% of the 63.5~kg of uranium underwent fission—yet this modest fraction still produced an explosive yield of approximately 15~kilotons TNT equivalent.

Meanwhile, the plutonium-production complex at Hanford, located along the Columbia River, reached full operation and generated substantial amounts of fissile plutonium. In April~1944, Emilio Segr\`e's group reported that reactor-produced plutonium contained a significant fraction ($\approx 18$--$20\%$) of $^{240}$Pu. Because $^{240}$Pu exhibits a high spontaneous-fission rate, its early-time neutrons would trigger pre-initiation in a gun-type assembly, leading to a fizzle. As a result, Los Alamos abandoned development of a plutonium gun-type weapon and turned to an alternative design.

Following Robert~Serber's Los Alamos lectures in April~1943~\cite{ser}, the fundamental principles of both gun-type and implosion-type assemblies were outlined, providing the theoretical foundation for later work. Motivated by these ideas, Seth Neddermeyer and collaborators began exploring the implosion concept~\cite{hod}. In this scheme, high-explosive charges are symmetrically arranged around a spherical core and detonated nearly simultaneously, generating an inward-directed shock wave that compresses the $^{239}$Pu pit to supercriticality.

Recognising its potential, Oppenheimer instructed the laboratory to pursue a fully spherical implosion design (see Fig.~\ref{fig8}, right). Achieving this goal proved extraordinarily challenging: it required development of explosive ``lenses'' to shape the detonation front and sub-microsecond timing circuits so that converging shock waves met precisely at the centre.

Even with these advances, the spontaneous-fission rate of $^{240}$Pu posed a continuing hazard. To reduce this risk, fuel elements in the Hanford reactors were irradiated for relatively short periods (typically 90--120 days)  before removal, keeping the $^{240}$Pu fraction below about 10\%, ideally closer to 7\%. Modern nuclear weapons—whether uranium-, plutonium-, or composite-based—ultimately trace their lineage to this implosion-type design.

A further innovation was the polonium--beryllium neutron initiator, later codenamed ``Urchin,'' developed at Los Alamos during 1944--45~\cite{reed3}. Polonium emits $\alpha$ particles that induce $(\alpha,n)$ reactions in beryllium, producing a sharp burst of neutrons at the instant of maximum compression, thereby ensuring dependable initiation of the chain reaction.

Behind these developments stood a formidable theoretical group led by Hans~Bethe. Among those regarded highly by Bethe was Klaus~Fuchs, a German-born British physicist who was later revealed to have passed classified information to the Soviet Union~\cite{clo}. His espionage, which dated from roughly 1943, facilitated the acceleration of the Soviet atomic programme in the late 1940s. Although the Soviet project had been reduced in priority after Germany's invasion in 1941, Stalin ordered its rapid expansion following the 1945 Potsdam Conference. Other security breaches within the Manhattan Project similarly contributed to the emergence of the post-war U.S.--Soviet nuclear rivalry.

Returning to the main narrative, the performance of the new implosion device could only be verified through full-scale testing. On 16~July~1945, the Trinity test—conducted in the New Mexico desert—confirmed the success of the design. The observed yield exceeded some theoretical predictions (e.g.\ those of I.~I.~Rabi) while falling short of others (notably Edward~Teller). As one of my professors, Emilio Segr\`e, later remarked,  ``I had never seen anything so beautiful in colour and shape.'' The implosion design demonstrated at Trinity formed the basis of the weapon detonated over Nagasaki.

\section{Political Aspects of 1945}\label{sec8}

The year 1945 marked several decisive political turning points. The sequence of major developments up to approximately July may be summarised as follows.

In February, prior to Germany’s surrender, the Allied Powers---the United Kingdom, the United States, and the Soviet Union---met at the Yalta Conference to discuss post-war arrangements, including the treatment of Germany and the future of Poland. Even before Germany’s capitulation, it was agreed that the country would be divided into zones of occupation under the joint administration of the Soviet bloc (East) and the Western Allies (West), with the capital, Berlin, similarly partitioned.

In addition, the secret Yalta Agreement on the Far East was concluded between the United States and the Soviet Union. It stipulated that two to three months after Germany’s defeat, the Soviet Union would enter the war against Japan, undertaking military operations in Manchuria, southern Sakhalin, and the Kuril Islands. Although Japan and the Soviet Union were still bound at the time by the Soviet--Japanese Neutrality Pact, this agreement was classified as a military secret and remained unknown to the Japanese government.

In April~1945, U.S.~President Franklin~D.~Roosevelt---whose health had been deteriorating---passed away on 12~April. On the same day, Vice~President Harry~S.~Truman succeeded him as President. Truman had only been briefly informed about the highly classified Manhattan Project. After Germany’s surrender, the war in the Pacific continued to rage, and it is difficult to imagine that, after such immense investment in the Manhattan Project, its momentum could have been halted.

Following the Yalta Conference, Soviet forces advanced into and ultimately occupied Berlin. On 30~April, shortly after Roosevelt’s death and as Soviet troops entered the city, Adolf~Hitler took his own life together with his wife, Eva~Braun. Consequently, Germany formally surrendered on 8~May (for Western Europe) and 9~May (for the Soviet Union), marking the end of World~War~II in Europe.

With Germany’s surrender in May~1945,the original and dominant motivation — fear of a possible German atomic bomb — had effectively disappeared. Nevertheless, many scientists, including Oppenheimer, continued to urge completion of the weapon. In fact, by late autumn~1944, when it had already become clear that Germany would not succeed in developing an atomic bomb, only a single scientist, Joseph Rotblat, formally resigned from the Manhattan Project \cite{and}.  On the other hand, up to the successful CP-1 experiment in~1942, nuclear research had been driven primarily by scientific curiosity. By~1945, however, the entire enterprise had come under strict military supervision.

\begin{figure}[htbp]
\centering
\includegraphics[width=1\textwidth]{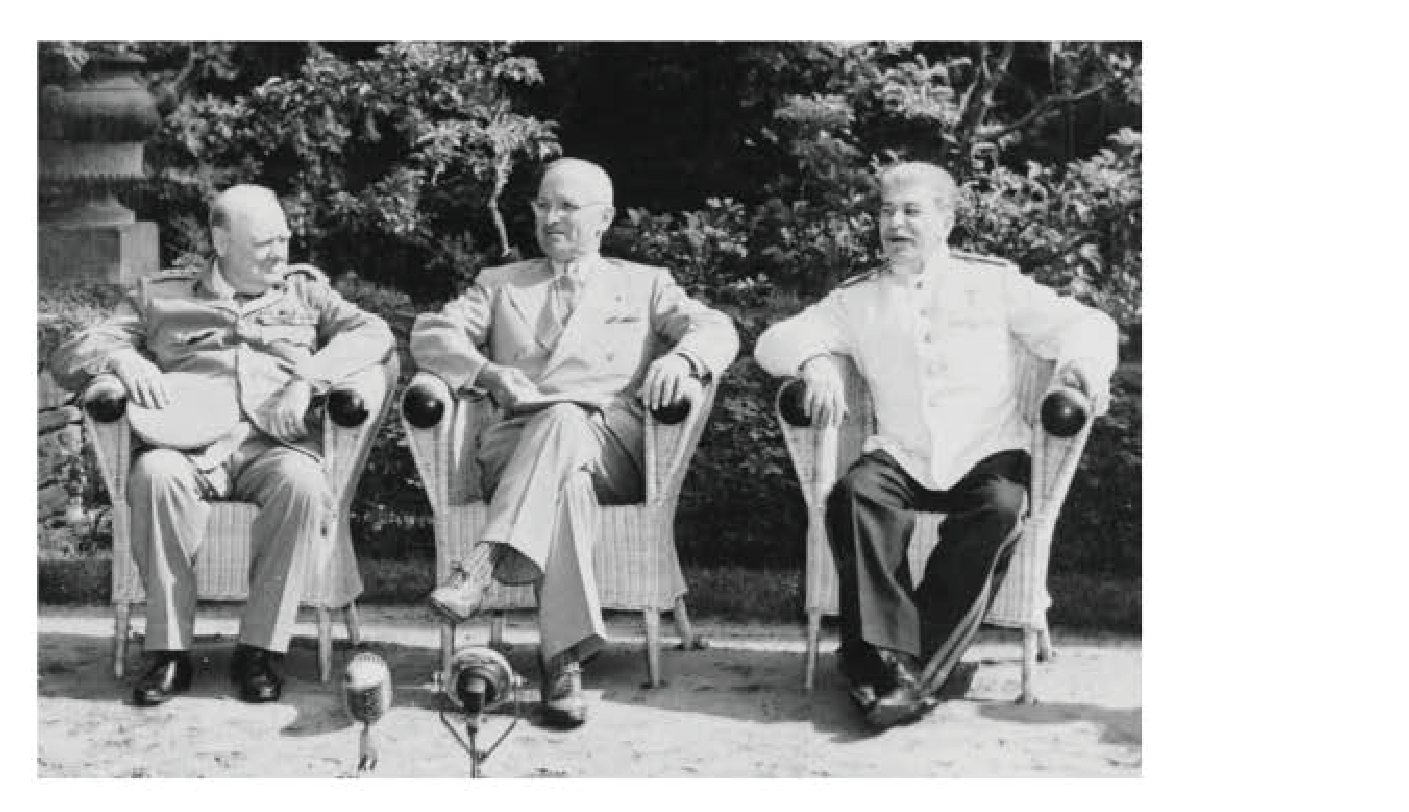}
\caption{The Potsdam Conference, which began on 17~July~1945. From left to right: Winston~Churchill (Prime~Minister of the~United~Kingdom), Harry~S.~Truman (President of the~United~States), and Joseph~Stalin (Premier of the~Soviet~Union).}
\label{fig9}
\end{figure}

From 17~July to 2~August~1945, the leaders of the United~Kingdom, the United~States, and the Soviet~Union met once again—this time in Potsdam, located in Soviet-occupied East~Germany—for the Potsdam~Conference (see Fig.~\ref{fig9}). 
Discussions centred on the post-war settlement of Europe and the Middle~East and culminated in the drafting of the Potsdam~Agreement. 
Meanwhile, just before the conference, on 16~July, the first nuclear test—codenamed \textit{Trinity}—had been successfully conducted in New~Mexico. 
In its aftermath, President~Truman began confidential consultations with Prime~Minister~Churchill. 
Under the terms of the 1943~Quebec~Agreement, the United~Kingdom and the United~States had pledged that the atomic bomb would not be used without mutual consent. 
Consequently, the decision to employ the weapon against Japan was reached through secret deliberations that excluded the Soviet~Union.

On 26~July~1945, the Potsdam~Declaration was issued, demanding Japan’s unconditional surrender. 
It was signed by the United~States, the United~Kingdom, and Chiang~Kai~Shek of China. 
Although Joseph~Stalin did not sign the declaration, he had already inferred that the United~States and the United~Kingdom were preparing to employ the atomic bomb against Japan. 
During the conference, President~Truman informed Stalin that the United~States possessed a new and powerful weapon that would soon be used against the Japanese, though he offered only a brief and deliberately vague remark.

The political manoeuvring between the Western~Allies and the Soviet~Union at this time was exceedingly complex, and from a specialist perspective, certain aspects of this narrative may remain incomplete or simplified. 
Nevertheless, 1945 was a year of extraordinary intensity, marked by rapid and far-reaching developments surrounding the atomic bomb.

Following the conference, Stalin returned to the Soviet~Union and began reinforcing the Soviet~Far-Eastern~Army in preparation for war with Japan. 
Thus, the first half of~1945 was characterised by a succession of major political developments outside Japan that decisively shaped the subsequent course of history.

\section{Before the Bombing of Japan and the Dilemma among Scientists}
\label{sec9}

Here, let us consider how the scientists involved in the Manhattan Project viewed the atomic bombing of Japan. Many of the British and American physicists were motivated to advance the development of the atomic bomb in order to oppose Nazi Germany. In particular, many of the Jewish scientists were driven by the desire to combat a regime that had subjected Jews to horrendous persecution. At that time, both the military and the scientific communities were working towards the same goal. However, after Germany’s surrender, it is likely that many of the scientists experienced more complex and conflicted emotions.

One example comes from the Metallurgical Laboratory at the University of Chicago, which played a key role in the Manhattan Project. At the request of General~Groves, a survey of scientists was conducted on 12~July~1945. About 40\% of the 150 respondents expressed opposition or hesitation about using the bomb against Japan \cite{wal}, revealing a growing sense of moral unease within the scientific community.

Among the members of the laboratory was James~Franck, best known for the ``Franck--Hertz experiment'', which laid an essential foundation for quantum theory. In~1934, he had been dismissed from his university position in Germany owing to the Nazi regime’s anti-Jewish policies and subsequently emigrated to the United~States via Copenhagen. On 11~June~1945, he became the principal author of the \textit{Franck~Report}~\cite{fra}, a document of nearly~5\,000~words that argued against the unannounced use of the atomic bomb on Japan.

Furthermore, on 17~July~1945, physicist Le\'o~Szil\'ard and around seventy other scientists involved in nuclear research raised their voices in protest~\cite{szi2}. A petition was sent to the President of the United~States, urging that the atomic bomb not be used without prior warning to Japan. Some of the signatories would later become Nobel~laureates. Below is an excerpt from the Japanese translation of that petition:

\begin{quote}
\textbf{Petition to the President of the United States (Excerpt)}\\[4pt]
We, the undersigned seventy scientists, have been engaged in research in the field of atomic energy. Until recently, we pursued this work during the war out of concern that Germany might use atomic bombs against the United States. Today, with Germany defeated, that threat has been removed. While the war must be brought to a swift and successful conclusion, and while an atomic attack on Germany might have served as an effective military measure, we feel strongly that such an attack on Japan cannot be justified unless the conditions being imposed on Japan are made public in full detail.
Such use of the atomic bomb should never be undertaken without serious consideration of the moral responsibilities involved. If, after this war, enemy forces are allowed to possess these new means of destruction without restraint, not only American cities but cities in other nations as well will face sudden and grave danger.
\textit{(Middle portion omitted.)}
The question of whether or not to use the atomic bomb, as stated in this petition, must be carefully considered in light of moral responsibility, and the final decision rests with you.
\end{quote}

Unfortunately, the petition was blocked by General~Groves and never reached the President. When I was at Berkeley, my friend David~Hendrie---who later became a senior official at the U.S.~Department of~Energy---told me that, after the bombing of Japan, about half of the young physicists of that generation left the field of physics altogether. This may, of course, be my own speculation, but while it is true that many supported the bombing, I suspect that a considerable number of scientists continued to harbour deep and conflicted feelings about it in their hearts.

\section{The View from the US Army toward Japan}\label{sec10}

From the American perspective, the key sequence of events between the outbreak of the Pacific War in December~1941 and the atomic bombings in August~1945 can be summarised as follows. The Battle of Midway in June~1942 resulted in a severe and unexpected defeat for the Imperial Japanese Navy, marking a major strategic turning point. However, U.S.\ military planners did not regard this as a sign that Japan might surrender soon; rather, they anticipated a long and costly campaign. In Japan, the extent of the defeat was not publicly acknowledged, as wartime propaganda suppressed details of the losses.

In the following years, numerous battles and large-scale bombing operations were conducted across the Pacific theatre. By 1945, the United States had begun to employ napalm on a massive scale in incendiary raids on Japanese cities. The gel, developed under Harvard chemist Louis~Fieser, produced devastating firestorms in densely built wooden urban areas. These attacks caused enormous civilian casualties through fire, radiant heat, and suffocation in the firestorm environment.

Even before Germany’s surrender, the U.S.\ military had begun identifying potential targets for the newly developed atomic bomb \cite{wel}. In its early deliberations, the Target Committee examined a list of seventeen cities. By late June~1945, the principal candidates had been narrowed to Kyoto, Hiroshima, Kokura, and Niigata---cities that had thus far avoided extensive conventional bombing and that were considered suitable for assessing the effects of the new weapon.

Shortly thereafter, however, U.S.\ Secretary of War Henry~L.\ Stimson---who had visited Kyoto twice and appreciated its cultural and historical importance---objected strongly to its inclusion as a target. Stimson insisted that Kyoto be removed from all bombing lists, including the atomic category. General Leslie~R.\ Groves, head of the Manhattan Project, repeatedly argued that Kyoto possessed valid military significance, but Stimson feared that its destruction would cause irreparable damage to post-war relations between Japan and the United States. President Truman ultimately accepted Stimson’s recommendation.

Thus, by the time of the Potsdam Conference in July~1945, the primary candidates for the atomic bombings were Hiroshima, Kokura, and Niigata, with Nagasaki added as a reserve target. On 6~August, the gun-type uranium bomb was dropped on Hiroshima. Three days later, on 9~August, the implosion-type plutonium bomb was dropped on Nagasaki. Kokura had originally been designated as the target for the second bombing, but adverse weather and heavy cloud cover forced the mission to divert to Nagasaki.

On 9~August, Japan notified the United States---via Switzerland and Sweden---that it was prepared to accept the Potsdam Declaration, provided that the Imperial institution be preserved. Although this represented a significant shift, U.S.\ leaders continued to insist on an unconditional surrender.

On 10~August, General Carl~Spaatz sent a message to General Lauris~Norstad, then responsible for target planning, indicating that additional atomic operations might be required if Japan did not accept the Allied terms. Norstad suggested that, should another strike become necessary, Tokyo---rather than Niigata---might be considered, owing to its remaining governmental and military significance. Some later secondary accounts have claimed that President Truman told the British ambassador on 14~August that, if Japan continued to resist, he might have ``no alternative'' but to order an atomic bomb to be dropped on Tokyo. As Wellerstein \cite{wel} has emphasised, however, no primary-source document confirms that such an instruction was formally issued or approved.

Before any further decision was made, Japan accepted the Allied demand for unconditional surrender on 15~August~1945, thereby ending the Second World War and precluding the use of a third atomic bomb.

Meanwhile, the Soviet Union---as stipulated in the secret Yalta Agreement, which called for Soviet entry into the Pacific War ``two to three months'' after Germany’s defeat---declared war on Japan on 8~August, exactly three months after Victory in Europe Day. Soviet forces launched their offensive the following day, advancing rapidly from Manchuria (then the puppet state of Manchukuo) into Korea, southern Sakhalin, and the Kuril Islands. Even after Japan’s announcement of surrender on 15~August, Soviet forces continued operations in Sakhalin and the Kurils. For this reason, the Soviet Victory Day in the Far East is commemorated on 3~September, a date that remains a point of sensitivity in Japan--Russia relations.

Had Japan not surrendered on 15~August, some American officials feared that continued Soviet advances might lead to the division of Japan after the war, possibly including the occupation of parts of Hokkaido. Indeed, Stalin expressed interest in establishing a Soviet occupation zone in northern Japan, but Truman rejected the proposal.

\section{Scientists in the Midst of the Bombing}\label{sec11}

As mentioned earlier, Ernest~O.~Lawrence invented the cyclotron in the early 1930s. In~1935, a few years later, Ryokichi~Sagane, then aged thirty, began working under Lawrence at the University of~California, Berkeley. About six years his junior, Luis~W.~Alvarez---an eclectic physicist later renowned for his hypothesis that an asteroid impact caused the extinction of the dinosaurs, for his wartime innovations in aircraft\hyp landing techniques, and for his 1968~Nobel~Prize in~Physics for contributions to particle physics---also joined the laboratory. The group collaborated harmoniously on a series of pioneering experiments.

In~1938, Sagane returned to Japan and resumed his professorship at the University of~Tokyo. He would later become vice-presient of the
Japan Atomic Energy Agency (JAEA). Toward the end of the war, just before the atomic bomb was dropped on Japan, Lawrence is reported to have remarked, ``I do not want to drop the bomb on a country where Sagane lives.'' It is uncertain whether Alvarez shared Lawrence’s sentiment, but he---together with Robert~Serber, who along with Oppenheimer would later be questioned during the Red~Scare, and Philip~Morrison, who subsequently became a prominent advocate of the anti\hyp nuclear\hyp weapon movement---sent a letter to~Sagane expressing their concern and friendship.

At that time, there was no direct means of communication, so the scientists decided to attach their letters to a bomb drop. Three instrument canisters, each containing a copy of the letter, were released over Nagasaki.  Fortunately, one of them, dropped on 9~August, was recovered by the Japanese military, as it had been attached to a parachute equipped with a short\hyp wave transmitter---although some accounts suggest that local citizens in the city of~Sasebo handed it over to the authorities.

\begin{figure}[htbp]
\centering
\includegraphics[width=0.85\textwidth]{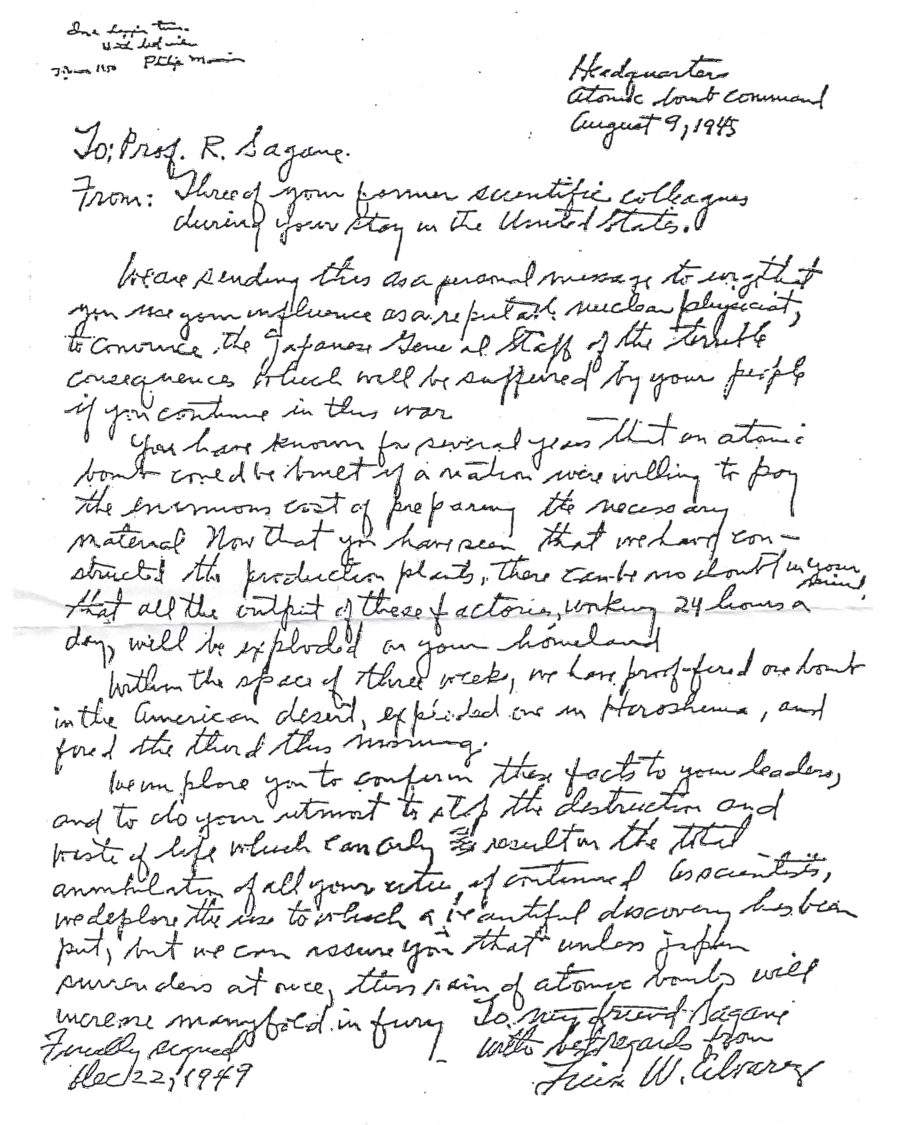}
\caption{Letter dropped with the Nagasaki bomb on 9~August~1945, addressed to Ryokichi~Sagane from Luis~Alvarez and two colleagues, although their names were intentionally omitted.}
\label{fig10}
\end{figure}

Some weeks after the bombing, Sagane was notified by the military that a letter addressed to him had been found. For the following two months, however, there was no further correspondence. Finally, in late~November or early~December~1945, he was contacted again and travelled to~Sasebo to retrieve the letter. It had been roughly folded and placed in a torn envelope. Sagane eagerly read the English letter on the spot in~Sasebo, as shown in Fig.~\ref{fig10}, and later translated it into Japanese~\cite{sag}. Because Sagane’s translation employed classical Japanese, I present below a modernised rendering for clarity and readability.

\begin{quote}
\textbf{Commander of the Atomic Bomb Command Headquarters}\\
\textbf{August 9, 1945}\\[4pt]
\textbf{To:} Professor Ryokichi~Sagane\\
\textbf{From:} Three of your colleagues in scientific research during your time in America\\[6pt]
Knowing that you are a distinguished nuclear physicist in Japan, we are writing to you in earnest. We believe that you fully understand the terrible consequences Japan will face if it continues this war. Please convey to the Imperial Army General Staff that prolonging the conflict will lead only to catastrophic results. It is with this heartfelt wish that we send this letter.
You have long been aware that, given sufficient national resources, an atomic bomb could be built. The United~States now possesses factories dedicated to this purpose, operating around the clock. It is therefore inevitable that the total output of these facilities will ultimately fall upon your homeland. Thus far, within three weeks, one bomb has been detonated in the American desert, one over Hiroshima, and this morning, a third has been dropped.
We earnestly implore you: if Japan continues the war, every city in the country will be destroyed. Please inform your nation’s leaders and do everything within your power to prevent further senseless loss of life and resources. \underline{As scientists, it pains us profoundly that such remarkable discoveries are being used in} \underline{this way}. However, if Japan does not immediately surrender, the rain of atomic bombs will grow even more intense. That, we fear, is now beyond doubt.
\end{quote}

Sagane repeatedly read and reread the underlined sections of the letter, recalling the friends with whom he had once shared a life of research. Eventually, he came to interpret the letter as follows:

\begin{quote}
\textit{``Was not our shared purpose in scientific research to contribute to the welfare of humanity? As your friends who once worked alongside you, we did not wish the atomic bomb to be used in this way. Now that you understand its destructive power, we urge you to take every possible action. That is the sincere wish of your three friends who once walked this path of science with you.''}
\end{quote}

As he travelled back to Tokyo from Sasebo, he was deep in thought. Although the senders were unnamed, Sagane sensed that one of them might have been his junior, Luis~Alvarez. Memories of Lawrence also came flooding back.

I first heard this story in Berkeley in the mid-1970s and shared it with my close friend at the time, Professor~Akira~Isoya of Kyushu~University. Isoya was then compiling a memorial volume in honour of his mentor, Ryokichi~Sagane, and was gathering related materials.

Shortly after returning to Japan, Isoya contacted Sagane’s daughter, Setsuko~Sengoku. By chance, Sengoku discovered the letter in a place where her father had once said, ``No one must ever touch this.'' It was during the thirteenth year memorial service. The full text of the letter was later published at the beginning of the \textit{Ryokichi~Sagane~Memorial~Volume}~\cite{sag}; in fact, it had also been submitted shortly after the war to a women’s magazine~\cite{sag}.

\begin{figure}[htbp]
\centering
\includegraphics[width=0.75\textwidth]{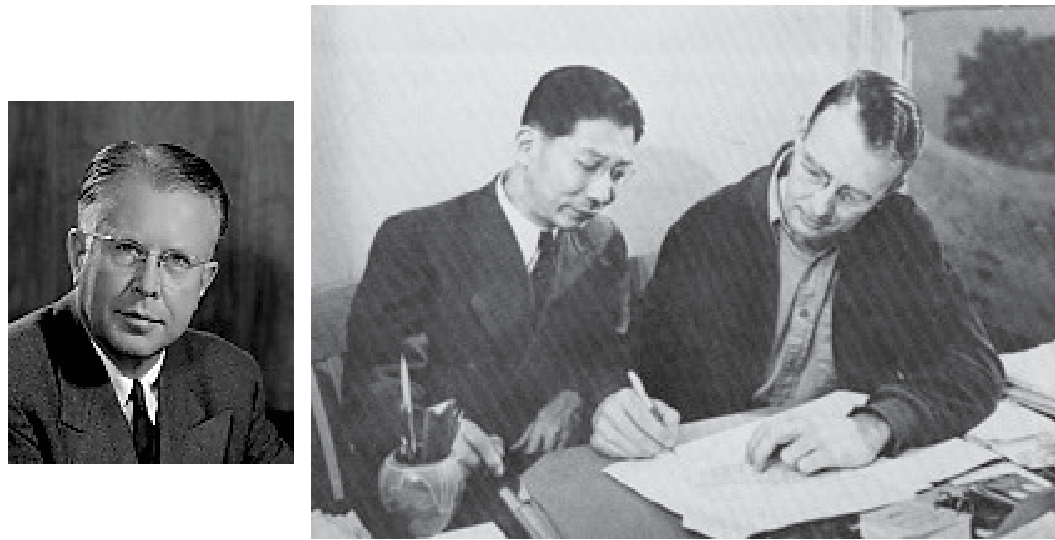} 
\caption{Ernest~O.~Lawrence, the supervisor of both Ryokichi~Sagane and Luis~Alvarez in the 1930s (left), and Sagane and Alvarez signing in December~1949 (right).}
\label{fig11}
\end{figure}

It was subsequently confirmed that the letter shown in Fig.~\ref{fig10} had been written primarily by Luis~Alvarez. On 22~December~1949, more than four years after the atomic bombing, Alvarez formally acknowledged and signed it. Figure~\ref{fig11} shows him signing the document. If one looks closely at the fine print in the lower left corner of Fig.~\ref{fig10}, this fact is apparent. By the end of~1949, the story had been picked up by an American newspaper and was also reported in the 3~January~1950 edition of the \textit{Yomiuri} newspaper in Japan.

Scientific research begins with curiosity. Yet whether it ultimately contributes to the happiness and welfare of humanity is a question that must always remain at the forefront of our minds. Through this episode---and others that I shall share next---I was once again deeply reminded of that truth. The following accounts, including this chapter and also Sections~12.2~(on~Chamberlain) and~13.3~(on~Ms~Cho), overlap with topics I have previously written about in Japanese~\cite{nag}, but I would like to revisit and share them here anew.

\section{Two Scientists After the War}\label{sec12}
\subsection{J. Robert Oppenheimer}

After the war, Oppenheimer came to believe that nuclear weapons represented a 
profound threat to humanity and could bring about its annihilation at any moment. 
This conviction led him to advocate nuclear disarmament and to work toward 
preventing a nuclear arms race with the Soviet Union. He was already aware that the 
Soviet Union, through espionage involving figures such as Klaus Fuchs, had embarked 
on its own nuclear weapons programme.

\begin{figure}[htbp]
\centering
\includegraphics[width=0.55\textwidth]{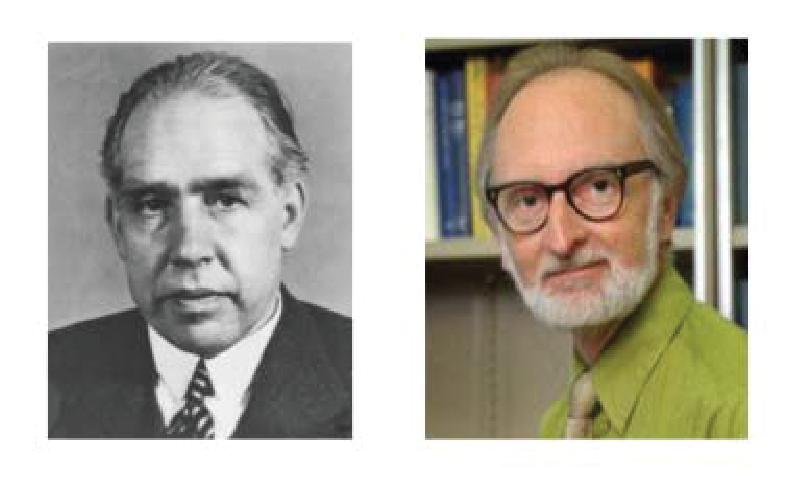}
\caption{Left: Niels~Bohr, whom Oppenheimer regarded as his greatest intellectual influence. Bohr also provided the United~States in~1939 with early insights into nuclear fission (see Chapter~4). Right: Owen~Chamberlain, who later became my supervisor at Berkeley and is discussed in the following subsection.}
\label{fig12}
\end{figure}

The person Oppenheimer most respected was Niels Bohr of Copenhagen (see Fig.~\ref{fig12}, left). Bohr held the firm belief that all scientific advances---whatever their 
nature or potential consequences---should be shared openly with the world. Because 
of this conviction, Winston Churchill regarded him with suspicion; nevertheless, 
Bohr never wavered in his principles.

Whether or not he was directly influenced by Bohr, Oppenheimer believed that the 
scientific principles underlying the atomic bomb would inevitably spread throughout 
the world. What truly mattered, he felt, was how to impose limits on its production---
a task he regarded as far more urgent. For instance, he opposed the rapid development of the hydrogen bomb during the 1949--1950 debates, particularly through the General Advisory Committee (GAC) report.

Meanwhile, against the backdrop of the U.S.--Soviet Cold War, several figures---led 
politically by Senator Joseph McCarthy and administratively by Atomic Energy 
Commission (AEC) Chairman Lewis Strauss---launched the so-called ``Red Scare,'' 
during which scientists were subjected to intense scrutiny and accusations of 
disloyalty. On 12 April 1954, the AEC revoked Oppenheimer's security clearance after a formal hearing, not for any proven breach of security but for alleged ideological unreliability.

Several researchers left Berkeley during that era. Although Oppenheimer himself had 
already departed Berkeley in 1943 to direct the Los Alamos Laboratory and later 
served as Director of the Institute for Advanced Study, he too became a central 
target of political attacks during this period. One of those who left Berkeley, 
Robert Serber, happened to have his office directly across the hall from mine when I 
was at Columbia University. I only exchanged brief greetings with him at the time, 
but looking back, I deeply regret not having learned more from him about that 
extraordinary era.

\subsection{Owen Chamberlain}

When I was around thirty and staying in Berkeley, I belonged to the Segr\`e--
Chamberlain group, where my direct supervisor was Owen Chamberlain (see the 
right-hand photograph in Fig.~\ref{fig12}). My first experiment in Berkeley began with a 
discussion I had with him while we were eating hamburgers for lunch. Even after I 
left Berkeley, he would often ask me to give a speech on his birthday or at his 
celebrations. He was also the first person to open my eyes to many of the themes 
discussed in this article---nuclear energy, the atomic bomb, and the Manhattan 
Project.

Shortly before his death, he sent me a personal postcard. By the time I received it 
in Japan, he had already passed away, sometime between February and March 2006. 
After his passing, I co-authored an obituary with Herb Steiner and John Jaros in 
\textit{Physics Today}~\cite{jar}.

Chamberlain had joined the Manhattan Project as a young student. He even gave me 
a book describing his experiences during that period. When I knew him, he was still 
actively conducting experiments, but he was also deeply engaged in social causes---%
especially as a leading voice in the anti--atomic-bomb movement. He advocated 
passionately for nuclear disarmament, appearing regularly on radio broadcasts. I 
often felt that he carried a profound sense of anguish over the atomic bombing of 
Japan. Yet, as seen in Fig.~\ref{fig12}, he always greeted me with a warm and gentle smile.

In 1988, while I was at Columbia University, he telephoned me. He told me that he 
had been invited to Japan by the publishing company \textit{Kodansha} and expressed a strong 
wish to visit Hiroshima and offer his apology. Although a visit to Hiroshima was 
not originally part of \textit{Kodansha}'s itinerary, he travelled there in March 1988 to 
express his remorse and lay flowers. His words---``The pain of those who were 
attacked comes back to me. We must not repeat our mistakes.''---were reported in 
the \textit{Chugoku Shimbun} newspaper~\cite{chugok}, and I understand that it was 
also covered by NHK television news at the time. Later, a serious article about him 
appeared in \textit{Sh\=onen Magazine}, published by Kodansha. I still keep a copy 
of that article.

From the time I first met him, Owen~Chamberlain was among the Americans who most profoundly regretted the atomic bombings.

\section{Survivors from the Atomic Bomb}\label{sec13}
\subsection{Inhabitants in New Mexico}

It is often assumed that the only victims of atomic bombs were those in Hiroshima 
and Nagasaki. While this is largely correct given the catastrophic death tolls in 
those two cities, the world's first atomic bomb had in fact been detonated earlier---%
on 16 July 1945---during the Trinity test in New Mexico. This raises the question 
of what became of the inhabitants who lived near the test site.

The Trinity site was selected on the premise that no permanent residents lived 
within a radius of approximately 20 miles (32~km). However, contemporaneous 
archival documents and numerous testimonies from local downwinders and their 
descendants indicate that this assumption was inaccurate. Many residents have 
reported adverse health effects, including thyroid disorders and various cancers, 
consistent with the well-documented biological consequences of radiation 
exposure~\cite{blu}.

The Trinity test was conducted under strict military secrecy, and residents of 
New Mexico received no prior warning. In the aftermath of the war, the United 
States established the Atomic Bomb Casualty Commission (ABCC) in Japan to 
investigate the long-term health effects on survivors of Hiroshima and Nagasaki. 
By contrast, no comparable organisation was ever created to study---or even to 
acknowledge---the victims of the Trinity test. The New Mexico Tumor Registry began 
to collect partial data only in 1966, nearly two decades after the explosion, and a 
comprehensive scientific analysis of the Trinity fallout plume did not appear in the 
peer-reviewed literature until 1987~\cite{cdc}.

Several studies have identified the five counties most heavily affected by 
radioactive fallout---Socorro, Lincoln, Guadalupe, San Miguel, and Torrance---%
together comprising a population of roughly 65,000 in the mid-1940s. A 2020 
investigation by the National Cancer Institute (NCI) concluded that these counties 
were at the greatest risk, estimating that several hundred excess cases of cancer, 
primarily thyroid cancer, were likely attributable to the Trinity detonation, with 
additional cases expected to emerge in subsequent decades~\cite{sim,lah}. The 
same report, however, noted that ``great uncertainty'' remains owing to the scarcity 
of contemporaneous meteorological measurements and the complete absence of 
systematic fallout monitoring in 1945.

\subsection{An Example of Hiroshima Residents}

The fourth and final episode brings the narrative closer to home---to my own 
extended family. While there is no need to recount once again the many horrifying 
accounts of survivors from Hiroshima and Nagasaki, I wish to share one story from 
among my relatives.

My father had an elder sister named Fumiko, my aunt. She married a man with the 
family name Ch\=o, becoming Fumiko Ch\=o, and they lived in Tokyo with their five 
children---a family of seven. Shortly before 1945, however, her husband's work 
transfer brought the family to Hiroshima.

\begin{figure}[b]
\centering
\includegraphics[width=0.7\textwidth]{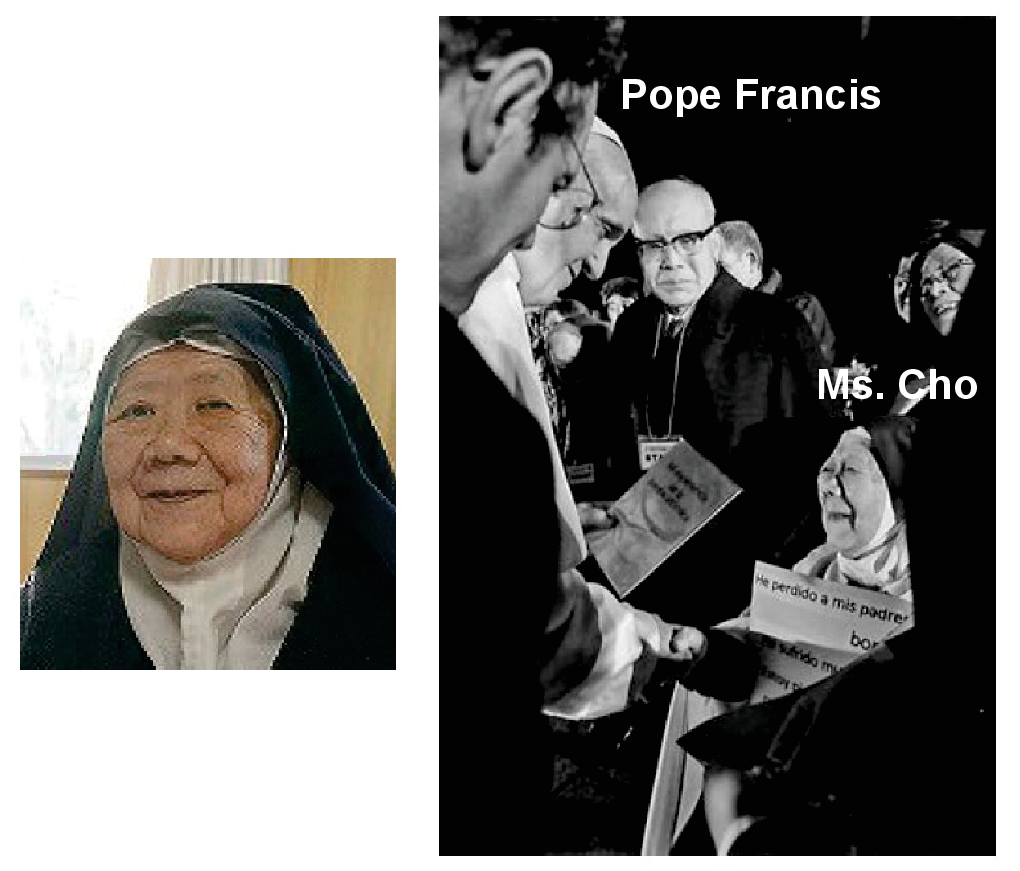}
\caption{Left: Yasuko~Ch\={o}. Right: Yasuko~Ch\={o} meeting Pope~Francis in November~2019.}
\label{fig13}
\end{figure}

Their second daughter, Yasuko Ch\=o, then fourteen years old---a sensitive age---%
appears in the left-hand photograph of Fig.~\ref{fig13}. The family's home stood 
approximately 3.5~km from the hypocentre. On the morning of 6 August 1945, with 
no school that day, Yasuko rose around eight o'clock. Everyone in the family except 
the youngest daughter, Saeko, was out in the city. After washing her face, she 
suddenly saw an intense flash of light. Moments later, the house collapsed, and she 
ran in panic to the air-raid shelter. The flash was the atomic bomb.

Yasuko's father, who was working downtown, was killed instantly. Her elder sister 
survived only a few hours, pleading for water before dying the next day. Her mother, 
terribly burned, managed to return home but succumbed soon afterward. Yasuko 
walked through several railway stations---through areas filled with the unbearable 
stench of death---searching for her younger sister, who had gone to school. All six 
hundred pupils had perished. Her elder brother, who had remained in Tokyo, 
survived; in Hiroshima, only Yasuko and Saeko were left alive. The memories of 
those days were so vivid and painful for Yasuko that even recalling them later in life 
filled her with terror.

At the age of twenty, Yasuko converted to Christianity and entered a convent. It 
was a strict order, yet given her experiences in Hiroshima, one can readily imagine 
that she found in religion the only possible path toward peace of mind. In later years 
she lived at the Carmelite Monastery in Yamaguchi City.

Around the year 2000, Junichi Yoshinaga---a playwright and her cousin (and mine 
as well)---urged her strongly to record her experiences, saying, ``If you do not leave 
a record, who will?'' With his encouragement, she eventually completed a written 
testimony in 2006~\cite{cho}. The account was so vivid and harrowing that I shall 
refrain from repeating its details here; suffice it to note that they were profoundly 
tragic.

Yet even amid such darkness, there was one ray of light. In November 2019, 
Pope Francis visited Hiroshima. Yasuko was chosen as one of the survivors to meet 
him. Before the meeting, her friends translated her testimony into Spanish and 
English and prepared a small placard summarising her words. When the Pope 
approached her, he gently placed his hands on her head, his face filled with sorrow 
(Fig.~\ref{fig13}, right). It was the first papal visit to Hiroshima in thirty-eight years, 
and for Yasuko---a devout Catholic nun---it must have been a moment of profound 
consolation and grace. The local newspaper covered the encounter extensively.

Returning to the main narrative: after the bombing, Yasuko and Saeko were taken 
in by my family in the small town of Sabae, Fukui Prefecture, through arrangements 
made by my father. Our own home in Mikage, near Kobe, had been completely 
destroyed by fire when I was about one year old in 1945, and we too had evacuated 
to Sabae. I later heard many stories about Yasuko and Saeko's time there from my 
elder brother, who, being nearly four years old at the time, remembered them vividly.

During those months, Yasuko often carried me on her back, and until quite 
recently she would still affectionately call me ``Sh\=o-chan, Sh\=o-chan.'' Yet I can 
only imagine the hardship she endured, living in a strange place so soon after losing 
her entire family. Eventually, the two sisters were sent to live with relatives near 
Yokohama.

At that time, tension among residents at the evacuation site, Sabae, was high. One local 
family protested, saying, ``We only agreed to take in four members of your family.'' 
Even in the countryside, economic hardship was severe. My father's parents also 
came to Sabae after losing their home in Tokyo. With eight people under one roof, my 
mother---pregnant at the time---was overcome by stress and passed away soon 
afterward, at about thirty years of age. I was only two and a half, and I retain no 
memory of her.

In 2023, Yasuko passed away at the age of ninety-two. Shortly afterward, I received 
a letter from her younger sister, Saeko, who wrote, ``I was so young at the time that I 
do not remember much about Hiroshima.'' Perhaps that very lack of memory spared 
her deeper anguish. As I read her words, I realised that, in a sense, the same may 
also be said of myself.

\section{Concluding Remarks}\label{sec14}

The narratives traced in this article focus on the brief yet transformative 
seven-year period from 1938---when Enrico Fermi left Italy for the United States 
and the discovery of nuclear fission was announced---to the atomic bombings of 
August~1945. During this brief interval, scientific and technological developments 
advanced from the identification of fission to the construction and military deployment 
of nuclear weapons at an extraordinary, perhaps unprecedented, pace. Even in retrospect, 
the rapidity of this progression remains striking. And precisely because of its speed, 
the consequences of these developments proved to be both profoundly beneficial and 
deeply destructive. Many of the scientists involved must have confronted intense and 
complicated emotions as events unfolded.

As noted earlier, scientific inquiry is invariably motivated by curiosity. Yet whether 
that curiosity ultimately contributes to the well-being of humanity is a question that 
must continually be examined. The history reviewed here underscores the necessity of 
such reflection.

Whenever I visit the Hiroshima Peace Memorial Museum, I am inevitably confronted by 
a complex set of emotions. Foremost among them is the memory of my cousins---the 
Ch\={o} sisters, Yasuko and Saeko---whose experiences defy easy description, as well as 
the memory of my mother, who suffered as a secondary victim of the war’s devastation. 
At the same time, I am reminded of the internal struggles of many American physicists 
who worked on the bomb and later grappled with its consequences. The image of Owen 
Chamberlain and his reflections after the war often returns to mind, as do the moral 
conflicts of others, including Le\'{o} Szil\'{a}rd, who carried deep misgivings. It is also worth 
recalling that many younger researchers left physics altogether in the aftermath of the 
Manhattan Project.

A realistic assessment of the context of the summer of 1945 must also acknowledge the 
strategic considerations of the time. While the destruction in Hiroshima and Nagasaki 
was horrific and the moral burden on scientists considerable, it is widely recognized 
that the bombings contributed to bringing the war to an end. A land invasion of Japan 
would likely have resulted in catastrophic casualties on all sides, and conventional bombing 
campaigns would almost certainly have continued. Furthermore, had Soviet forces participated 
in an invasion, Japan might have emerged from the war as a divided nation, with far-reaching 
consequences for the subsequent course of history.

Yet war inflicts damage not only on those who suffer directly, but also on those who---
whether intentionally or not---cause harm through their actions. As Hibakusha organizations 
and many moral leaders worldwide have emphasized, war itself, along with the possession and 
use of nuclear weapons, must never be accepted as inevitable. Having been born during the war, 
I feel this lesson with particular force: scientific progress must always be accompanied by an 
ethical awareness of its potential consequences for humanity.

By revisiting these interwoven scientific and human experiences, this article seeks to 
contribute to a broader historical understanding of how physicists confronted---and were 
themselves transformed by---the moral dimensions of their discoveries. It is my hope that 
such reflections may help inform the ways in which scientific research is guided in the future.

\section*{Acknowledgements}

Finally, I wish to express my deepest gratitude to the many individuals mentioned in this article with whom I had the privilege of speaking. Professors Emilio~Segr\`e, Glenn~Seaborg, Owen~Chamberlain, and~T.~D.~Lee graciously invited us to dinner on numerous occasions. I also had the opportunity to exchange words, albeit briefly, with Edward~McMillan, Luis~Alvarez, I.~I.~Rabi, Robert~Serber, and~Edward~Teller. In reflecting upon these towering figures, I have found renewed inspiration to write this article.

I am also deeply grateful to Professor~Ryoji~Noyori, former Director of RIKEN, whose encouragement during a lunch conversation in May~2024 inspired me to undertake this work, and to Professor~Georg~Wolschin of the University of Heidelberg for his kind support in publishing this article in English.


\begin{thebibliography}{99}

\bibitem{nev}
A.~Nevala-Lee, \textit{Collisions: A Physicist's Journey from Hiroshima to the Death of the Dinosaurs} 
(W.~W.~Norton \& Company, New York, 2025).

\bibitem{seg}
D.~L.~Heiserman, ``Element 43: Technetium,'' in \textit{Exploring Chemical Elements and Their Compounds} 
(New York, 1992), p.~164;  
E.~Segr\`e and G.~T.~Seaborg, ``Nuclear Isomerism in Element 43,'' \textit{Phys. Rev.} \textbf{54}, 772 (1938).

\bibitem{fermi}
E.~Fermi, ``Artificial Radioactivity Produced by Neutron Bombardment,'' Nobel Lecture, Stockholm, 12~December~1938, 
available at \url{https://www.nobelprize.org/prizes/physics/1938/fermi/lecture/}.

\bibitem{fermi1}
E.~Fermi, \textit{Enrico Fermi: His Work and Legacy} (Springer, Berlin, 1972).  
Available at: \url{https://link.springer.com/book/10.1007/978-3-662-01160-7}.

\bibitem{mcm}
E.~M.~McMillan and P.~H.~Abelson, ``Radioactive Element 93,'' \textit{Phys. Rev.} \textbf{57}, 1185 (1940).

\bibitem{sea}
G.~T.~Seaborg, E.~M.~McMillan, J.~W.~Kennedy, and A.~C.~Wohl, 
``Plutonium, the Element of Atomic Number 94,'' \textit{Phys. Rev.} \textbf{69}, 366 (1946).

\bibitem{stu}
R.~Stuewer, \textit{The Age of Innocence: Nuclear Physics between the First and Second World Wars} 
(Oxford University Press, Oxford, 2018).

\bibitem{lee}
T.~D.~Lee, \textit{Selected Papers and Letters of G. B. Pegram} (World Scientific, 1992).

\bibitem{reed1}
B.~C.~Reed, ``An Inter-Country Comparison of Nuclear Pile Development during World War II,'' 
\textit{Eur. Phys. J. H} \textbf{46}, 15 (2021).

\bibitem{reed2}
B.~C.~Reed, \textit{The Frisch--Peierls Memorandum: The Founding Document of the Nuclear Age} 
(Springer, Cham, 2022).

\bibitem{tube}
NHK BS Documentary \textit{Codenamed Tube Alloys: The Atomic Bombing and Churchill’s Strategy}, 
Parts~I \& II (Tokyo, 2019, in Japanese.).

\bibitem{wein}
A.~M.~Weinberg~(ed.), \textit{The Collected Works of Eugene Paul Wigner}, Vol.~5 
(Springer--Verlag, New York, 1992).

\bibitem{rho}
R.~Rhodes, \textit{The Making of the Atomic Bomb} 
(Touchstone, New York, 1986).

\bibitem{reed3}
B.~C.~Reed, \textit{Manhattan Project: The Story of the Century} 
(Springer, Cham, 2020).

\bibitem{ser}
R.~Serber, \textit{Los Alamos Lectures on the Manhattan Project} (Los Alamos, 1943).

\bibitem{hod}
L. Hoddeson \textit{et~al.}, \textit{Critical Assembly: A Technical History of Los Alamos during the Oppenheimer Years, 1943–1945} 
(Cambridge University Press, 1993).

\bibitem{clo}
F.~Close, \textit{Trinity: The Treachery and Pursuit of the Most Dangerous Spy in History} 
(Penguin Books, London, 2020).

\bibitem{and}
A. Brown, \textit{The Keeper of the Nuclear Conscience: The Life and Work of Joseph Rotblat} (Oxford University Press, 2012).

\bibitem{wal}
J. S. Walker, \textit{Prompt and Utter Destruction: Truman and the Use of Atomic Bombs Against Japan}  (University of North Carolina Press, 1997, rev. ed. 2004)

\bibitem{fra}
J.~Franck \textit{et~al.}, \textit{Report of the Committee on Political and Social Problems}, 
Metallurgical Laboratory, University of Chicago (1945).  
Available at: \url{https://ahf.nuclearmuseum.org/ahf/key-documents/franck-report/}.

\bibitem{szi2}
L.~Szilard \textit{et~al.}, \textit{Petition to the President of the United States}, 
Metallurgical Laboratory, University of Chicago (1945).  
Available at: \url{https://ahf.nuclearmuseum.org/ahf/key-documents/szilard-petition/}.

\bibitem{wel}
A.~Wellerstein, ``Did the U.S. Plan to Drop More Than Two Atomic Bombs on Japan?'' 
\textit{National Geographic}, 4~August~2020.  
Available at: \url{https://www.nationalgeographic.com/history/magazine/2020/07-08/did-united-states-plan-drop-more-than-two-atomic-bombs-japan/}.

\bibitem{sag}
R.~Sagane, \textit{Memorial Collection of Ryokichi Sagane} (privately published, 1981); originally written in 1946\,(in~Japanese).

\bibitem{nag}
S.~Nagamiya, \textit{Kagaku} (Iwanami Shoten, Tokyo) \textbf{90}, 1114 (2020)\,(in~Japanese).

\bibitem{jar}
J.~Jaros, S.~Nagamiya, and H.~Steiner, ``Owen Chamberlain,'' \textit{Phys. Today} \textbf{59}(8), 70 (2006).

\bibitem{chugok}
\textit{Chugoku Shimbun Newspaper} (1988), ``Newspaper Article about Owen Chamberlain'' (in~Japanese).  
Available at: \url{http://www.hiroshimapeacemedia.jp/?p=25914}.

\bibitem{blu}
L.~M.~M.~Blume, ``Collateral Damage: American Civilian Survivors of the 1945 Trinity Test,'' 
\textit{Bull. Atom. Sci.}, 17~July~(2023).  
Available at: \url{https://thebulletin.org/premium/2023-07/collateral-damage-american-civilian-survivors-of-the-1945-trinity-test/}.

\bibitem{cdc}
Division of Cancer Epidemiology and Genetics, U.S.~National~Cancer~Institute, ``Trinity Test: Community Summary'' (2023).
Available at: \url{https://dceg.cancer.gov/research/how-we-study/exposure-assessment/trinity/community-summary}.

\bibitem{sim}
S.~L.~Simon, ``Estimated Radiation Doses and Cancer Risks Resulting from Exposure to Fallout from the Trinity Nuclear Test,'' 
\textit{Health Phys.} \textbf{119}, 389--397 (2020).

\bibitem{lah}
Los~Alamos~Historical~Document~Retrieval~and~Assessment~Project, 
\textit{Final Report of the Los Alamos Historical Document Retrieval and Assessment Project},  
prepared for the Centers for Disease Control and Prevention (November~2010), pp.~ES-34--35.
Available at: \url{https://nsarchive.gwu.edu/document/33249-document-34-center-disease-control-and-prevention-national-center-environmental}.

\bibitem{cho}
Y.~Cho, \textit{Atomic Bomb Testimony} (Yamaguchi Carmelite Monastery, 2006).  
(Japanese, English, and Spanish editions.)

\end{thebibliography}
\end{document}